\documentclass{JHEP3}

\usepackage{graphicx}
\usepackage{dcolumn}
\usepackage{bm}

\def\be{\begin{equation}}
\def\ee{\end{equation}}
\def\ba{\begin{eqnarray}}
\def\ea{\end{eqnarray}}

\newcommand{\bea}{\begin{eqnarray}}
\newcommand{\eea}{\end{eqnarray}}

\def\nn{\nonumber}

\def\a{\alpha }
\def\eps{\epsilon}
\def\p{\partial}

\def\T{ {\cal T}}
\def\W{ {\cal W}}

\def\ZZ{{\mathbb Z}}
\def\w{\omega }
\def\td{\tilde }

\preprint{UB-ECM-PF-09/01}

\title{On Schwinger Pair Creation in Gravity and in Closed Superstring Theory}

\author{Jorge G. Russo\\Instituci\' o Catalana de Recerca i Estudis Avan\c{c}ats (ICREA),\\
Institute of Cosmos Sciences and Department ECM, University of Barcelona, Av.Diagonal 647,  Barcelona 08028 SPAIN}

\date{\today}

\abstract{
We investigate  the Schwinger pair creation process in the context of gravitational models with the back reaction
of the electric field included in the geometry. The background is also an exact solution of type II superstring theory, where
the electric field arises by Kaluza-Klein reduction. 
We obtain a closed formula for the pair creation rate that incorporates the gravitational back reaction.
At weak fields it has the same structure as the general Schwinger formula, albeit pairs are produced by
a combination of Schwinger and Unruh effect, the latter due to the presence of a Rindler horizon.
In four spacetime dimensions, the rate becomes constant at strong electric fields.
For states with mass of Kaluza-Klein origin, the rate has a power-like dependence in the electric field,
rather than the familiar (non-perturbative)  exponential dependence. 
We also reproduce the same formula from the string partition function for winding string states.
Finally, we comment on the generalization to excited string states.
 }

\keywords{Schwinger effect, Gravity, String theory}

\begin{document}

\section{Introduction}

The  Schwinger effect of
pair creation in the presence of an electric field 
is represented by a probability  given by the formula \cite{Schwinger}
\be
{\cal W}= T V_{d-1} {2j+1\over (2\pi )^{d-1}} \sum_{k=1}^\infty (-1)^{(2j+1)(k+1)} \left({eE\over k}\right)^{d\over 2} \ e^{-{\pi k M^2\over |eE|}}
\label{uno}
\ee
where $E$ is the electric field and $e,j,M $ are the charge, spin and mass of the pair created particles, and $d$ is the number
of uncompact spacetime dimensions.

The phenomenon has been shown to hold in open superstring theory \cite{bachas}. In this case, there is no gravity at tree level and the uniform electric field with flat spacetime geometry
is a consistent  background of the theory. 
In $d$ uncompact dimensions,
one finds a  result which at weak fields approaches 
the Schwinger formula (\ref{uno}) for the infinite collection of particles of the open superstring spectrum.

An important question is whether the phenomenon persists in the presence of gravity, where the back reaction to electric
field configurations can play an important role.
There are no static uniform electromagnetic field configurations in gravity. The basic reason is that electromagnetic flux lines attract each other
and tend to cluster around some central region.
Examples of axially symmetric, but non-uniform magnetic field
configurations in closed (bosonic and super) string theory have been constructed and investigated in \cite{exactly,RT2}.
In particular,  
there is a two-parameter $(b,\tilde b)$ class of models given by \cite{exactly,RT2}
$$
ds^2= -dx_0^2+dr ^2+  {r ^2\over 1+\tilde b^2r ^2} \big( d\varphi + (b+\td b)  d y \big) \big( d\varphi + (b-\td b)  d y \big) + dy^2+ dx_1^2+...+dx_{D-4}^2 \ ,
\nn
$$
\be
e^{-2(\phi -\phi_0)}=1+\tilde b^2r ^2\ ,\qquad   B_2=\tilde b {r ^2\over 1+\tilde b^2r ^2} d\varphi\wedge d y\ ,
\label{kkm}
\ee
Here $ y$ is a periodic coordinate, $ y= y+2\pi R$,  $D=10$ for type II superstring and $D=26$ for closed bosonic string theory.
More generally, this background solves the equations of the action
\be
S_D=\int d^{D}x \sqrt{G}\ e^{-2\phi}\left( R+4 (\p_\mu \phi )^2 -{1\over 12} (H_{\mu\nu\rho})^2\right)
\label{GenR}
\ee
in any spacetime dimensions.

Dimensional reduction in the $ y$ direction gives two magnetic fields
of different $U(1)$'s, proportional to $b$ and $\tilde b$, associated with $g_{y\varphi}$ and $B_{y\varphi}$, respectively. T-duality exchanges the parameters $b$ and $\tilde b$.
Setting the parameter $\tilde b$ to zero, one finds the Kaluza-Klein (KK) Melvin solution found in  \cite{GM} and subsequently investigated in \cite{Dowker3} in the context of gravitational field theories.
The background (\ref{kkm}) is an exact solution of (super)string theory to all $\alpha' $ orders and
the corresponding string conformal sigma model can be solved exactly \cite{exactly,RT2}. In particular, the full physical string spectrum can be obtained in terms of creation and annihilation operators, much like in the free string case. 

As pointed out in \cite{exactly}, and investigated in \cite{Costa,Pioline,Berkooz,Verlinde} for the model with $\tilde b=0$, electric field configurations can be obtained from (\ref{kkm}) by a  Wick rotation in the coordinates $x_0\to ix_{D-3}$ and $\varphi\to -i t $,
so that $t $ now plays the role of time.
By setting the parameter $\tilde b$ to zero and changing $b\to i E$ one finds the electric version of the Kaluza-Klein Melvin model,
which we shall call the $E$-model
\be
E:\ ds^2=dr ^2 -  r ^2 (dt - E d y)^2 +d y^2+  dx_i dx_i\ ,\ \ \ i=1,...,D-3\ ,
\label{Amodel}
\ee
with constant dilaton and vanishing $B_2$-field. The space is locally flat, but it has non-trivial identifications.

The T-dual solution, that we shall call $\tilde E$-model, is obtained from (\ref{kkm}) by setting $b=0$ and  changing $\tilde b\to i \tilde E$, 
\bea
\tilde E :\ &&ds^2= dr ^2 -  {r ^2\over 1 - \tilde E^2r ^2} dt^2 + {dy^2 \over 1 - \tilde E^2r ^2} + dx_i dx_i\ ,\ 
\nn\\
&&e^{-2(\phi -\phi_0)}=1 -\tilde E^2r ^2\ ,\qquad   B_2=\tilde E { r ^2\over 1-\tilde E^2 r ^2} dt\wedge d y\ ,
\label{Bmodel}
\eea
The geometry is singular at $ r  =1/\tilde E$. Nonetheless, the string model is  regular, as follows from the fact that T-duality
gives an equivalent conformal field theory and the $E$-model (\ref{Amodel}) is obviously regular.

String states are characterized by their (integer) winding number $m$ and KK momentum $n$ in the compact direction $ y$ as well as by quantum numbers
representing internal excitations.  The T-duality that maps the $E$-model  to the $\tilde E$-model  exchanges $n$ and $m$, so that a given string state of winding and KK momentum charges $(m,n)$  
behaves in the $E$-model  in exactly the same way as a string state with charges $(n,m)$ (changing, at the same time, $R\to \a'/R$). 
In any of the two models, the presence of either charge, $m$ or $n$ produces
a non-trivial interaction with the electric field.

The geometry (\ref{Amodel})  was investigated in the context of General Relativity in \cite{Verlinde}, where no Schwinger pair production  for KK particles was found
(though it was found some particle creation of different origin).
The interpretation of this fact was that the back reaction of geometry prevents the electrostatic potential from overcoming the rest mass of the KK particles, preventing the tunneling
that would otherwise give rise to the Schwinger effect. An interesting question is whether this is a model-dependent result or we should expect that
gravitational back reaction will always prevent the standard Schwinger phenomenon to take place.

The present results clarify this point. In consistency with \cite{Verlinde}, we will show that in the $E$-model  indeed there cannot be any Schwinger pair production of KK particles.
However, we find that there is Schwinger pair production
for {\it winding} string states. Such states cannot be investigated within the context of field theory,
but one can consider the $\tilde E$-model, where these winding states become KK particles. Our results imply that in General Relativity coupled to a dilaton and an antisymmetric tensor $B_2$
as in (\ref{GenR}) there is Schwinger pair production of KK particles in the background (\ref{Bmodel}), as we shall discuss below. 
In this case we recall that the electric field arises
from the $B_{ty}$ component.

\medskip 

This paper is organized as follows. 

In section 2 we  review the Schwinger effect for charged scalar particles, first 
in the more familiar case of an electric field
in Minkowski spacetime \cite{Schwinger}
(see \cite{padma1,padma2,page} for more recent discussions); then for an electric field in Rindler spacetime \cite{rindler}.

In section 3 we compute the pair creation rate for a charged scalar field in the $\td E$- model (\ref{Bmodel}).
We find that the probability for pair creation is given by 
\be
 {\cal W} ={{\cal T}V_{d-2}\over (2\pi)^{d-1} |\td E|} \sum_{k=1}^\infty {(-1)^{k+1}\over k} \left( {2|q\td E|\over k} \right)^{d-1\over 2} e^{-{\pi k M_0^2\over 2|q\td E|}}   \ 
{\rm erf}(Y\sqrt{k}) \ ,\qquad 
\label{primera}
\ee
$$
Y\equiv  \sqrt{ {\pi  |q \td E| \over 2 \td E^2} }\ ,
$$
where ${\rm erf}(z)$ is the standard error function and $q=n/R$ is the Kaluza-Klein charge of the state. $M_0$  is the mass in the higher $D$ dimensional spacetime
and the mass in $d\equiv D-1$ dimensions is $M^2=M_0^2+q^2$. 
$\W $ exhibits a number of interesting physical features that we discuss.

In section 4 we discuss string theory  in the background (\ref{Amodel}). In section 4.1, we first reproduce the formula of section 3
 for (unexcited) winding string states. Then, in section 4.2, we comment on some issues 
to derive a formula for general excited winding string states.

\section{Charged scalar  field }

The solution (\ref{Bmodel}) represents an electric field on a space which approaches a Rindler space at small $ r $. 
Before considering this space, 
it is useful to recall some features of the dynamics of charged particles coupled to an electric field in Minkowski and in Rindler spacetime.

\subsection{Electric field in Minkowski spacetime}

To study  Schwinger pair creation in flat Minkowski spacetime, $ds^2=-dt^2+dz^2+dx_i^2$, $i=1,...,d-2$,
we choose a gauge $A_t=  E z$ and consider the equation for
a massive charged scalar particle minimally coupled to the electromagnetic field, 
\be
\big( -\p_z^2-\p_i^2 + ( \p_t -i e E z )^2 +M^2 \big) \Phi =0
\ee
Setting $\Phi =e^{ip_i x_i+i\w t}\chi(z)$ one gets the Schr\" odinger equation for a particle in an inverted harmonic potential $V=-(eEz-\w )^2$,
\be
\big[ - \partial_z^2 -(eEz- \w )^2 + p_i^2 +M^2\big]\chi (z) =0\ ,\qquad 
\label{flatE}
\ee
There are many different derivations of the Schwinger rate (\ref{uno}). In this first example we 
will compute the pair production rate from the partition function for the magnetic model obtained by analytic continuation, $t=i x$, $B=iE$, $\w=-i p$, $x_{d-2}=-ix_0$, $p_{d-2}=i p_0$.
This gives a harmonic oscillator with Hamiltonian
\be
H= - \partial_z^2 +(eBz-p)^2 + p_j^2-p_0^2 +M^2\ ,\qquad j=1,...,d-3
\ee
and  eigenvalues  (we take $eB>0$)
\be
H_n = 2eB (n+{1\over 2}) + p_j^2-p_0^2 +M^2\ ,\qquad n=0,1,2,...
\ee
This leads to a one-loop vacuum energy 
\be
Z = \int_0^\infty  {ds\over s} \ {\rm Tr}\big[ e^{-\pi sH}\big] 
%
\label{xagar}
\ee
The trace contains a summation over the momenta, which is converted into an integral by the rule
\be
\sum_{p_j,p_0,p}\to    {V_{d-1}\over (2\pi )^{d-1} } \int d^{d-3}p_j \ dp_0\ dp 
\ee
The integral over $dp$ gives a $\delta(0)$ which is interpreted as a time volume factor (recall that $-ip$ is the energy in the electric configuration). This is  
just what is needed to get a finite probability per unit time; 
since $dp$ has the same dimensions as $eB dt$, the integral over $dp$ gives a factor $eBT$.
Thus
\be
Z = {TV_{d-1}\over (2\pi)^{d-1} }\int_0^\infty  {ds\over s^{d\over 2}} \ {eB \over \sinh (\pi eB s)}\ e^{-\pi M^2 s}
\ee
In doing the analytic continuation back to the original electric field configuration,  an infinite series of poles arise from $\sin(\pi eEs)=0$, i.e. at $s=k/(eE)$, $ k=1,2,...$.
  which give rise to an imaginary part, given by $\pi $ times the residue of the poles (the integration contour passes by the right of the poles). 
This  represents the pair production probability
We get
 \be
{\cal W}_{\rm scalar}=2{\rm Im}Z= {TV_{d-1}\over (2\pi)^{d-1}}\sum_{k=1}^\infty (-1)^{k+1 }
 \left({|e E| \over k}\right)^{d\over 2} \ e^{-{\pi k M^2\over |eE|}}
\label{scalarrate}
\ee
in agreement with (\ref{uno}) with $j=0$. This is equivalent to the Lorentzian approach based on the 
Schwinger representation of the Feynman propagator by means of the kernel.




Another equivalent way to compute the pair production rate  is to choose the gauge where the gauge potential is $A_3=-Ex_0$. The equation
of motion is then similar, changing $z$ by $x_0$, but now we can interpret pair production in terms of a scattering process, where an ``in" wave hits the inverted harmonic potential 
and as a result there is a combination of negative and positive frequency waves at late times,  where the coefficients are determined by a Bogoliubov transformation.
The coefficient of the negative frequency component gives the pair creation probability for a given frequency. The similar calculation in the static gauge will be carried out
 in section 3.


\subsection{Electric field in Rindler space}

The case of Rindler space is obtained as follows.
We first consider a charged scalar particle moving in a uniform magnetic field in flat spacetime, but now we use radial coordinates and choose the gauge 
where $A_\varphi = B r^2/2$
(i.e. $A_x=-B y/2,\ A_y=Bx/2 $). 
The equation is now given by 
\be
\Big( {1\over r} \p_r  r \p_r  + {1\over r^2} \big( \p_\varphi -{i\over 2}\ e Br^2 \big)^2 +\p_j^2-\p_0^2  - M^2 \Big) \Phi =0\ ,
\ee
where $j=1,...,d-3$.
Setting $\Phi= e^{i p_0 x_0+i l \varphi +ip_j x_j}\chi(r)/\sqrt{r}$, we obtain
\be
\big[ - \partial_r^2 + V(r) \big]\chi (r) =0\ ,\qquad V(r)=  \left( {1\over 2} eBr - {l\over r} \right)^2-{1\over 4r^2} +M^2+p_j^2-p_0^2
\label{Rindler}
\ee
This represents a two-dimensional oscillator. The eigenvalues are 
\bea
H &=& eB (l_L+l_R+1) -eB (l_L-l_R) +M^2+p_j^2-p_0^2\ ,
\nn\\
&=& eB (2l_R+1)  +M^2+p_j^2-p_0^2\ ,\qquad l_L, \ l_R=0,1,2,...
\eea
where the quantum numbers are related to the Landau level $l$ and to the radial quantum number $k_r$ by
\be
l=l_L-l_R\ ,\qquad l_L+l_R =2k_r +|l|\ ,\qquad k_r=0,1,2,...
\ee
One can then compute the partition function and analytically continue back to the electric field configuration,
which is now obtained by analytic continuation $B=iE,\ \varphi=-it,\  x_0=i x_{d-2}$ and $l=i\w $, giving the Rindler space
$$
ds^2= dr^2-r^2 dt^2+dx_i^2 \ , \qquad i=1,...,d-2\ .
$$  
Although the starting ``uniform magnetic field" configuration is the same as in sect. 2.1, the analytic continuation
is  different:  now it is the polar angle what becomes the time coordinate, giving as a result a Rindler space, with a Rindler horizon located
at $r=0$. As shown in \cite{rindler}, in Rindler space one again obtains  the Schwinger pair creation rate, modulo a term that scales as the area
and represents Unruh particle production from the Rindler horizon.
The rate can be obtained by a Bogoliubov transformation between creation/annihilation operators
associated with in and out vacua at $r=0$ and $r=\infty $.
Generalizing the result of \cite{rindler} to $d$ dimensions, the formula for the rate (re-obtained in the next section) is given by 
\bea
{\cal W} &=& {{\cal T}V_{d-2}\over (2\pi)^{d-1}} \int d^{d-2}p \int_{-\infty}^0 d\omega\ \log {1+ e^{-{\pi (M^2+p_i^2)\over eE}} \over 1+ e^{-{\pi (M^2+p_i^2)\over eE}} e^{2\pi \w} }\ 
\nn\\ 
&=& {\cal W}_1 - {\cal W}_2
\label{gabo}
\eea 
where 
\bea
{\cal W}_1 &=&  {{\cal T}V_{d-2}\over (2\pi )^{d-1}}\int d^{d-2}p \int_{-\infty}^0 d\omega\  \log \big( 1+ e^{-{\pi (M^2+p_i^2)\over eE}} \big)
\nn\\ 
{\cal W}_2 &=&  {{\cal T}V_{d-2}\over (2\pi )^{d-1}}\int d^{d-2}p \int_{-\infty}^0 d\omega\ \log \big(  1+ e^{-{\pi (M^2+p_i^2)\over eE}} e^{2\pi \w } \big)
\label{gabr}
\eea 
We assume $eE>0$. 
Expanding the log and performing the integrals, one finds
\be
{\cal W}_1 = {V_RV_{d-2}\over (2\pi)^{d-1}} \sum_{k=1}^\infty (-1)^{k+1} \left( {eE\over k} \right)^{d\over 2} e^{-{\pi k M^2\over eE}}
\label{adomi}
\ee
We used the relation \cite{rindler} ${\cal T} d\w =eE dV_R$, where $V_R$ is the volume in the two-dimensional Rindler spacetime. 
This can be regularized by two limiting hyperbolas at $r_1$ and $r_2$.
Then one can write $V_R= T (r_2-r_1)$, where $T$ is a mean proper time, $T=\T \bar r$, $\bar r=(r_1+r_2)/2$ representing a typical distance
from the Rindler horizon (typical values are given by  the turning points of classical trajectories in the WKB approximation).
Thus ${\cal W}_1 $ represents the standard Schwinger probability for pair creation proportional to the spacetime volume $V_{d-2} V_R$.

The second contribution is a surface contribution, which has to be subtracted to the dominant contribution (\ref{adomi}) proportional to the volume.
The integral over $\omega $ is now convergent, giving the result
 \be
{\cal W}_2  = {{\cal T}V_{d-2}\over (2\pi)^d} \int d^{d-2}p  \sum_{k=1}^\infty {(-1)^{k+1}\over k^2} e^{-{\pi k (M^2+p_i^2)\over eE} }
\label{Li}
\ee
Integrating over $p_i$, we find
 \bea
{\cal W}_2  &=&  {{\cal T}V_{d-2}\over (2\pi )^d}  \sum_{k=1}^\infty {(-1)^{k+1}\over k^{2}} \left( {eE\over k} \right)^{{d\over 2}-1} 
e^{-{\pi k M^2\over eE} }
\nn\\
&=& - {{\cal T}V_{d-2}\over (2\pi )^d}  \big( eE \big)^{{d\over 2}-1} {\rm Li}_{1+{d\over 2}}\big( - e^{-{\pi  M^2\over eE}} \big)  
\label{Lia}
\eea
This second contribution is due to the presence of the Rindler horizon \cite{rindler}. 
In the absence of electric fields, this term plays the role of canceling the particle production in the Boulware vacuum state.
When the electric field is turned on, one term becomes proportional to the volume and the surface term becomes significant only 
in the vicinity of the horizon, where it is getting most of the contribution.







\section{Electric field in a gravitational theory
}

Thus far we have considered electric fields in fixed (Minkowski or Rindler) backgrounds, and neglected the back reaction of the geometry due
to the energy density provided by the electric field. We will now incorporate this back reaction exactly by using the string model 
(\ref{Bmodel}).

We shall  first consider  a massless scalar supergravity mode $\Phi $ 
in the background (\ref{kkm})
 This  satisfies the equation\footnote{The same equation was considered in \cite{magnetic} in the study of the mass spectrum of the magnetic model.}
\be
\partial_\mu ( e^{-2\phi }\sqrt{G} G^{\mu\nu}\partial_\nu )\Phi = 0\ .
\label{jala}
\ee
Using eq. (\ref{kkm}) we obtain
\bea
&&\big[-\p_0^2 +\p_k^2+{1\over r}\p_r(r\p_r) +{1\over r^2}(1+b^2 r^2) (1+\tilde b^2 r^2)\p_\varphi^2
\nn\\
&&+(1+\tilde b^2 r^2)\p_y^2- 2b (1+\tilde b^2 r^2) \p_\varphi \p_y\big] \Phi =0\ ,
\label{miq}
\eea
with $k=1,...,D-4$.
Write
\be
\Phi = e ^{ip_0 x_0+i p_k x_k+ i q  y + i l \varphi } {1\over \sqrt{r}}\ \eta (r) \ ,\qquad q = {n\over R}\ ,\ \ n\in \ZZ\ .
\ee
Then eq. (\ref{miq}) becomes
\be
\big[ - \partial_r^2 + V(r) \big]\eta (r) =0\ ,\qquad V(r)= {l^2-{1\over 4}\over r^2}+\nu^2 r^2+\mu^2\ ,
\label{mia}
\ee
where
$$
\nu = \tilde b  (q- bl)\ ,\qquad \mu^2 = p_k^2 -p_0^2+(q-bl)^2 +\tilde b^2 l^2\ .
$$
This is a two-dimensional oscillator with frequency $\nu $.
We thus obtain essentially the same differential equation as in the Rindler case, (\ref{Rindler}), but with different parameters. 

This mode $\Phi$ has  (Kaluza-Klein) momentum charge but vanishing winding number (winding states cannot be described by local fields).
Upon analytic continuation to the electric field configuration
one finds 
an inverted harmonic oscillator potential $-|\nu |^2r^2$, which is
again the origin of the instability that leads to Schwinger pair production.
However, note that 
this term is absent in the $E$-model (\ref{Amodel}), where $\tilde E=0$, since $\nu=0$ in this case. This implies that no Schwinger pair production of Kaluza-Klein particles should be expected in this model.
This explains the results of \cite{Verlinde}, which considered KK particles in the $E$-model (\ref{Amodel}).

We now return to the magnetic variables and set $b=0$. Therefore 
we will examine pair production of Kaluza-Klein particles in the $\tilde E$-model. 
Then we get
$$
\nu=  q \tilde b  \ ,\qquad  \mu^2 = p_k^2-p_0^2 +M^2+\tilde b^2 l^2\ ,\qquad M=|q|\ .
$$
Since we started with the  equation (\ref{jala}) for a {\it massless} scalar fluctuation,  the invariant mass of the state in the lower dimensional theory in the absence of an electric field coincides with the electric charge of the field, $M=|q|$. Below we shall  also consider massive states.

Now the Hamiltonian eigenvalues are
\be
H= 2 q\tilde b  (l_L+l_R+1) +\tilde b^2l^2 +M^2+p_k^2-p_0^2\ ,\qquad l_L, \ l_R=0,1,2,...,
\label{jamil}
\ee
$$
M^2\equiv q^2={n^2\over R^2}\ ,\qquad l=l_L-l_R,\qquad q\tilde b >0
$$
Comparing with the previous Minkowski and Rindler cases, the Hamiltonian  now contains a new term $\tilde b^2 l^2$.
The origin of this term is the $\tilde b^2  r ^2$ term in the metric component $g^{\varphi\varphi}=(1+\tilde b^2 r^2)/r^2 $. 
This term $\tilde b^2 l^2$ is produced by the back reaction of the magnetic field in the geometry.

\subsection{Calculation of the pair-production rate}

Consider now the analytic continuation to the electric field configuration, $\tilde b\to i\tilde E$ and $l\to i\w$, where $\w $ is the Rindler energy.
The calculation of vacuum persistence rate is formally the same as in \cite{rindler}, since the equation is the same with
the substitutions
\be
eE\to 2 q\td E \ ,\qquad   M^2\to M^2+ \td E^2\omega^2 -  2q \td E \omega
\label{spind}
\ee
It will be seen below that the new term $\td E^2\omega^2$ 
has a dramatic effect in the pair creation process.
As mentioned above, this term originates from gravitational back reaction, whereas the second term is to cancel the term $-eB l\to 2q\td E\omega $
in eq. (\ref{Rindler}), since this term does not appear in the differential equation (\ref{mia}).
The absence of such term $2q\td E \omega$ in (\ref{mia}) seems to be due to the Kaluza-Klein nature of the interaction.

The problem  can be viewed as the standard quantum mechanical problem of scattering
against a barrier (see e.g. \cite{padma1,padma2,page}). The pair creation rate will be related to the reflection  coefficient.
The starting point is the differential equation satisfied by the 
wave function $\psi(r) = \eta(r)/\sqrt{r}$:
\be
\psi '' +{1\over r}\ \psi '+ \big( q^2 \td E^2\ r^2+ {\w^2\over r^2} - \mu^2\big) \psi =0
\ee
with
\be
\mu^2=M^2+p_i^2+\w^2 \td E^2\ ,\qquad M^2=q^2\ ,\qquad i=1,...,D-3
\ee
(whereas $\mu^2=M^2+p_i^2+eE\w $ for the previous Rindler case).
It is important to note that, despite the classical singularity of the background at $r=1/\td E$, the
wave equation is regular at this point and can be extrapolated beyond this radius.

It is convenient to introduce new variables (throughout we assume $qE>0$)
\be
z=  q \td E\ r^2\ ,\qquad
\psi = 
z^{ {i\over 2}\w }\ e^{-i{z\over 2}}\ g(z)\ .
\ee
Then $g(z)$ satisfies the differential equation
\be
z\ g''(z)+ (1+i \w - iz)\ g'(z) -\a \ g(z)=0\ ,
\ee
with
\be
\a \equiv {\mu^2\over 4 q \td E}- {\w\over 2} +{i\over 2}\  .
\ee
The solution is given in terms of the confluent hypergeometric function
\be
g(z) =  A \ z^{-i\w}  \ {}_1F_1(-i\a -i\w ,\ 1- i \w ;\ iz)+ B \ {}_1F_1(-i\a ,\ 1+ i \w ;\ iz) \ .
\ee
Near $z=0$, $\psi(z)$ behaves as
\be
\psi (z) \cong   A\ z^{ -{i\over 2}\w }
+B\ z^{ {i\over 2}\w } \ ,\ \ \ z\sim 0\ .
\ee
Defining $z=e^{2x}$, one can write the $z=0$ behavior in terms of incoming and outgoing ordinary plane waves,
\be
e^{i\w t}\psi (x) \cong  A\ e^{ i\w (t -x)  }
+B\  e^{ i\w (t +x) } \ ,\ \ \ x\to -\infty\ .
\ee
At $z\to \infty $, $\psi$ has the following behavior
\be
\psi (z) \cong C\ z^{i\a + {i\over 2}\w} \ e^{-i{z\over 2}} + D\ z^{-i\a -1- {i\over 2}\w } e^{i{z\over 2}}\ ,\ \ \ \ z\gg 1\ ,
\ee
where 
\bea
C &=& e^{\pi\a\over 2} \left({B \ \Gamma(1+i\w ) \over \Gamma(1+i\w +i\a )} + {A\ e^{\pi\w\over 2} \Gamma(1-i\w ) \over \Gamma(1+i\a )}\right)
\nn\\
D &=& e^{\pi(\a-i)\over 2} \left( {B\ e^{\pi\w\over 2}\Gamma(1+i\w ) \over \Gamma(-i\a )} + {A\ \Gamma(1-i\w ) \over \Gamma(-i\a -i\w )}\right)
\eea
Therefore
\be
e^{i\w t} \psi( z) \cong {1\over\sqrt{z}}\ e^{i\w t}\left(  C\ z^{ {i\mu ^2 \over 4q\td E}} \ e^{-{iz\over 2}} + D\ z^{-{i\mu ^2 \over 4q\td E} } e^{{iz\over 2}} \right) \ .
\ee
We must demand that there is no wave coming from infinity. For positive frequency modes, this is the condition 
 $D\equiv 0$. This gives 
\be
\left| {B\over A} \right|^2 =\left| { e^{-{\pi \w \over 2 }}  \Gamma(-i\a )\over \Gamma(-i \a -i\w  )} \right|^2 =  {1+ e^{-{\pi (M^2+p_i^2)\over 2q\td E}} e^{-\pi ( {\w^2 \td E^2\over 2q \td E} + \w ) }\over 1+ e^{-{\pi (M^2+p_i^2)\over 2q\td E}} e^{-\pi ({\w^2 \td E^2\over 2q \td E} - \w ) } } < 1
\ee
For negative frequency, the condition is $C\equiv 0$, and we get an equivalent result:
\be
\left| {B\over A} \right|^2 =\left| { e^{{\pi \w \over 2 }}  \Gamma(1+ i\w+ i\a )\over \Gamma(1 +i \a  )} \right|^2 =  {1+ e^{-{\pi (M^2+p_i^2)\over 2q\td E}} e^{-\pi ( {\w^2 \td E^2\over 2q \td E} - \w ) }\over 1+ e^{-{\pi (M^2+p_i^2)\over 2q\td E}} e^{-\pi ({\w^2 \td E^2\over 2q \td E} + \w ) } } < 1
\label{rifle}
\ee
The Rindler space, as well as the present spacetime (\ref{Bmodel}), can be  divided in four quadrants, left, right, future and past.
As shown in \cite{rindler}, the sum $2\sum_{\w>0}+2\sum_{\w<0}$ gives the total volume. 
Quantization in the right quadrant corresponds to keeping only $\w < 0$.
If we were to integrate over both regions,  $\w>0$ and $\w < 0$, we would get an extra factor of 2.

Equation (\ref{rifle}) gives the reflection probability  of a given mode, which 
is equivalent to the probability for vacuum persistence
or vacuum to vacuum transition, 
\be
\big| \langle 0,\ {\rm out}\ \big| \ 0, \ {\rm in}\rangle \big|^2 = \left| {B\over A} \right|^2= \exp\Big( -2 VT \ {\rm Im} {\cal L}_{\rm eff} \Big)
\ee
where ${\cal L}_{\rm eff} $ is the effective Lagrangian. The pair-production probability is then\footnote{
Note that ${\cal W} \equiv  2 VT \ {\rm Im} {\cal L}_{\rm eff}$ has the interpretation of pair-production probability 
only to leading order in the semiclassical approximation, while eq.(3.20) is an exact relation (for a discussion see \cite{gavrilov}). 
We thank S. Gavrilov for emphasizing this point.}
\be
{\cal W} = 2 VT \ {\rm Im} {\cal L}_{\rm eff} = - \log      \left| {B\over A} \right|^2 
\ee
Integrating over the momenta and over the frequency,  the full pair creation probability is  given by 
\be
{\cal W} = {{\cal T}V_{d-2}\over (2\pi )^{d-1}}  \int d^{d-2}p \int_{-\infty}^0 d\omega\ \log {1+ e^{-{\pi (M^2+p_i^2)\over 2q\td E}} e^{-\pi ( {\w^2 \td E^2\over 2q \td E} + \w ) }\over 1+ e^{-{\pi (M^2+p_i^2)\over 2q\td E}} e^{-\pi ({\w^2 \td E^2\over 2q \td E} - \w ) } }
\label{fara}
\ee
where $d=D-1$ is the number of uncompact spacetime dimensions.
This exactly reproduces  the result (\ref{gabr}) of the Rindler model 
upon the formal substitution (\ref{spind}), as expected. 

For the present, gravitational case, the physical picture now involves important  differences, having to do with the presence of the factor $e^{-{\pi  \w^2 \td E^2\over 2q \td E}}$,
induced by a gravitational correction $\Delta M^2= \w^2 E^2 $ to the mass squared  $M^2=q^2= n^2/R^2$ of the Kaluza-Klein particle.
Expanding the logarithm and integrating over $p_i$, we now find
\be
 {\cal W} = {\cal W}_1 -{\cal W}_2
\ee
with
\bea
{\cal W}_1 &=& 
{{\cal T}V_{d-2}\over (2\pi)^{d-1} } 
\sum_{k=1}^\infty {(-1)^{k+1}\over 2q\td E} \left( {2q\td E\over k} \right)^{d\over 2} e^{-{\pi k M^2\over 2q\td E}}    
\int_{-\infty}^0 d\omega\ e^{-\pi k ({\w^2 \td E^2\over 2q \td E} + \w )} \ ,
\nn\\
{\cal W}_2 &=& {{\cal T}V_{d-2}\over (2\pi)^{d-1} }  \sum_{k=1}^\infty {(-1)^{k+1}\over 2q\td E} \left( {2q\td E\over k} \right)^{d\over 2} 
e^{-{\pi k M^2\over 2q\td E}} \int_{-\infty}^0 d\omega\ e^{-\pi k ({\w^2 \td E^2\over 2q \td E} - \w )   }
\ .
\label{derek}
\eea
Integrating over $\w $, we obtain
\bea
{\cal W}_1 &=& 
{{\cal T}V_{d-2}\over 2(2\pi)^{d-1} \td E} \sum_{k=1}^\infty {(-1)^{k+1}\over k} \left( {2q\td E\over k} \right)^{d-1\over 2} 
e^{-{\pi k (M^2-q^2)\over 2q\td E}}    
\big( 1+ {\rm erf}(Y \sqrt{k})\big) \ ,
\nn\\
{\cal W}_2 &=& {{\cal T}V_{d-2}\over 2(2\pi)^{d-1} \td E}  \sum_{k=1}^\infty {(-1)^{k+1} \over k}\left( {2q\td E\over k} \right)^{d-1\over 2} 
e^{-{\pi k (M^2-q^2)\over 2q\td E}}   
\big( 1- {\rm erf}(Y \sqrt{k} )\big) \ ,
\eea
$$
Y\equiv  \sqrt{ {\pi  q \td E \over 2 \td E^2} }\ ,
$$
where ${\rm erf}(z)$ is the standard error function.
Remarkably, the integrals over $\w $ are convergent: there is no infinite volume factor coming from the radial coordinate.
Both $\W_1 $ and $\W_2$ are proportional to $\T V_{d-2}$, just like the surface term $\W_2 $ in the Rindler case of section 2.2.
Near the horizon $\T\sim T/\bar r$, where  as in the Rindler case $T$ is proper time and $\bar r$ is a characteristic distance, so the pair production rate $\W/T$ goes roughly like $1/\bar r$, becoming more important in the vicinity of the horizon.



%

Since in the present case $M^2=q^2$, there is an exact cancellation of two exponential factors appearing in ${\cal W}_1$ and ${\cal W}_2$.
Therefore
\bea
{\cal W}_1 &=& 
{{\cal T}V_{d-2}\over 2(2\pi)^{d-1} \td E} \sum_{k=1}^\infty {(-1)^{k+1}\over k} \left( {2q\td E\over k} \right)^{d-1\over 2} 
\big( 1+ {\rm erf}(Y \sqrt{k})\big) \ ,\
\label{huy}
\\
{\cal W}_2 &=& {{\cal T}V_{d-2}\over 2(2\pi)^{d-1} \td E}  \sum_{k=1}^\infty {(-1)^{k+1} \over k}\left( {2q\td E\over k} \right)^{d-1\over 2} 
\big( 1- {\rm erf}(Y\sqrt{k})\big) \ .
\label{hut}
\eea

Now consider particles with $M^2>q^2$.
Our starting point is a scalar field with action
\be
S= \int d^{D}x \sqrt{G}\ e^{-2\phi}\left( G^{\mu\nu} \p_\mu \Phi \p_\nu \Phi + M_0^2 \Phi^2 \right)\ .
\ee
The equation of motion is given by
\be
\partial_\mu ( e^{-2\phi }\sqrt{G} G^{\mu\nu}\partial_\nu )\Phi = \sqrt{G}e^{-2\phi } \ M_0^2 \Phi\ .
\label{jama}
\ee
For the background (\ref{kkm}), the differential equation is the same as (\ref{miq}) with a term $M_0^2\Phi$ on the right hand side.
The subsequent equations are the same with the substitution of $M^2$ by $M^2\equiv M_0^2+q^2$.
Thus, proceeding in the same way as above, the general pair creation rate is then
\be
 {\cal W} ={{\cal T}V_{d-2}\over (2\pi)^{d-1} \td E} \sum_{k=1}^\infty {(-1)^{k+1}\over k} \left( {2q\td E\over k} \right)^{d-1\over 2} e^{-{\pi k M_0^2\over 2q\td E}}   \ {\rm erf}(Y\sqrt{k}) \ 
\label{floro}
\ee
This is the main result of this paper.

\subsection{General properties}

The sums over $k$ are convergent for both $\W_1$ and $\W_2$. This is seen from the behavior of the error function ${\rm erf}(z)$: it is monotonically increasing
with a linear behavior near $z\sim 0$ and tending to 1 at  $z\gg 1$. 
The asymptotic expansion of the error function is given by
\be
\sqrt{\pi }\ e^{z^2} \big( 1-{\rm erf}(z)\big) =\sum_{n=0}^\infty {(-1)^n\over 2^n z^{2n+1} } \ (2n-1)!!
\ee
Therefore, for any fixed electric field and $k\gg 1 $, we can substitute $ {\rm erf}(Y\sqrt{k})\to 1$ and the series (\ref{floro}) 
is readily seen to be convergent for the cases of interest, viz. $d\geq 2$ (including the case $M_0=0$, eqs. (\ref{huy}), (\ref{hut})).

\medskip

For weak electric fields, $Y$ is large, and we find
\bea
\big( 1+ {\rm erf}(Y\sqrt{k})\big) & \cong &  2+ O\big( e^{-kY^2} \big)
\nn\\
\big( 1- {\rm erf}(Y\sqrt{k})\big) & \cong & {1\over \sqrt{\pi k}\ Y}\ e^{- {\pi k q \td E\over 2\td E^2}}
\label{palas}
\eea
In this limit ${\cal W}_2 $ becomes negligible. The dominant contribution to the rate is
\be
{\cal W} \cong  {\cal W}_1 \cong 
{{\cal T}V_{d-2}\over (2\pi)^{d-1} \td E} \sum_{k=1}^\infty {(-1)^{k+1}\over k} \left( {2q\td E\over k} \right)^{d-1\over 2} e^{-{\pi k (M^2-q^2)
\over 2q\td E}} 
\label{fuerte}
\ee
Thus, at weak fields, the probability for pair creation for particles with $M>|q|$ has the same structure as in the Schwinger formula (\ref{uno}). More precisely, one
has a relation of the form 
$\W_{\rm grav}^{(d)}(M^2_0) ={\rm const.} {1\over 2q\td E^2 } \W_{\rm schw}^{(d+1)}(M^2_0)$.

Interestingly, for the case $M=|q|$, the rate is not exponentially suppressed. We obtain
\bea
{\cal W} &\cong & {{\cal T}V_{d-2}\over (2\pi)^{d-1} \td E} \sum_{k=1}^\infty {(-1)^{k+1}\over k} \left( {2q\td E\over k} \right)^{d-1\over 2} 
\nn\\
  &\cong & {{\cal T}V_{d-2}\over (2\pi)^{d-1} \td E} c_0\ (2q \td E)^{d-1\over 2} \ ,\qquad c_0=(1-2^{1-d\over 2})\zeta({d\over 2}+{1\over 2})
\label{greed}
\eea
 This seems to be related to  the fact that such state is massless in the original higher dimensional geometry.

Now let us  examine the behavior of the pair creation rate at strong electric fields. In this limit, $Y$ is small.
The argument of the error function is small for $k \ll k_0 \sim  O(E)$. Since the series is convergent, terms with $k>k_0$ will give a subleading contribution.
For the terms with $k \ll k_0 $ that provide the leading contribution, one has
\be
{\rm erf}(Y\sqrt{k})  =  {2\over \sqrt{\pi }}\ Y \sqrt{k} +O(Y^3)\cong   \sqrt{ {2 k q \td E \over  \td E^2} } 
\ee
Therefore we find
\bea
{\cal W}&  \cong &
{{\cal T}V_{d-2}\over (2\pi)^{d-1}\td E} \sum_{k=1}^\infty {(-1)^{k+1}\over k} \left( {2q\td E\over k} \right)^{d-1\over 2} 
\sqrt{ {2 k q \td E \over  \td E^2} }  \
\nn\\
& \cong & {{\cal T}V_{d-2}\over (2\pi)^{d-1}\td E^2} c_0'\  (2q \td E)^{d\over 2} \ ,\qquad c_0'=(1-2^{1-{d\over 2}})\zeta({d\over 2})\ ,\ \ d\neq 2\ ,
\eea
and $c_0'=\log(2)$ for $d=2$. Note that this is the leading behavior  also for particles with $M>|q|$.
In the particularly interesting case of $d=4$ the rate goes to a constant in the strong field limit:
\be
\W_{d=4}\bigg|_{\td E  = \infty}  = {\cal T}V_{2}\ {q^2\over 24\pi } \ .
\ee

Comparing the weak and strong field behaviors of $M=|q| $ particles, for weak fields  we have $\W\sim E^{d-3\over 2}$ whereas for strong fields
$\W\sim E^{d-4\over 2}$. Therefore, for these particles, the strong field behavior in $d$ dimensions is the same as the weak field behavior in $d'=d-1$.

\medskip 

Summarizing, the behavior of $\W $ for particles with $M=|q|$ is as follows. 
\smallskip

\noindent $\bullet $ For $d=2$ $\W $ is divergent as $\td E\to 0$ (see (\ref{greed})) and goes to zero like $1/\td E$ as $\td E\to \infty $. 
The divergence at zero field may look puzzling, since for weak fields,  gravitational back reaction is small and 
one would expect to recover results similar to those of the model of  
sect. 2.2, which is regular as $E\to 0$, even for particles with $M=0$. 
This will be clarified in the next section.

 \smallskip

\noindent $\bullet $ For $d=3$,  $\W $ has a finite, non-vanishing value at $\td E=0$, given by
$\W = \T V_{1}\ q/24$. As $\td E$ is increased, the rate decreases monotonically and, as $\td E\to \infty $,
 it goes to zero like $1/\sqrt{\td E}$ .
\smallskip
 
\noindent $\bullet $ For $d=4$, $\W $ is monotonically increasing, beginning from 0 at $\td E=0$ and approaching a constant 
${\cal T}V_{2}\ q^2/( 24\pi )$ as $\td E\to \infty $.

\smallskip
 
\noindent $\bullet $ For $d>4$, $\W $ is monotonically increasing, beginning from 0 at $\td E=0$ and going to infinity as $\td E\to \infty $.
 
 \smallskip

{}For particles with $M>|q|$, the leading behavior at $\td E=\infty $ is the same as that of particles with $M=|q|$. 
The weak field behavior, given in eq. (\ref{fuerte}), is quite different, instead. 
As $\td E\to 0$, $\W $ vanishes  exponentially as $\exp(-\pi M_0^2/(2qE))$  in all dimensions. 
In particular, this shows that for $d=2,3$, the rate  has a maximum 
at some finite value of the electric field (since for these values of $d$,  $\W$ vanishes also at $\td E=\infty $).

\subsection{ Discussion}

To  connect with the Rindler case of sect. 2.2, we have to consider a situation
where the gravitational correction $\Delta M^2= \w^2 \td E^2$ can be neglected.
In particular, it should be much smaller than the 
mass squared $M^2=M_0^2+ q^2$, which is the case for modes with frequency $\w^2 \ll M^2/\td E^2$.
Therefore we shall come back to eq. (\ref{fara})
and  integrate over $\w $ with a cutoff at some $\w_0^2 \ll M^2/\td E^2$. Expanding the log and integrating over $p_i$, we have (see eq. (\ref{derek}))
\be
{\cal W} = {{\cal T}V_{d-2}\over (2\pi)^{d-1}} \sum_{k=1}^\infty  {(-1)^{k+1}\over 2qE } \left( {2q\td E\over k} \right)^{d\over 2}
\int_{-|\w_0|}^0 d\omega\  
 e^{-{\pi k M^2\over 2q\td E}} e^{-{\pi k\w^2 \td E^2\over 2q \td E} } \big( e^{-\pi k\w}-e^{\pi k\w} \big) 
\label{jara}
\ee
Define $\epsilon \equiv \w_0 \td E/M $. Then for $\w\sim \w_0$ the exponent in (\ref{jara}) is\footnote{Here we assume both $q>0$ and $\td E>0$.
The cases of $q<0$ and/or $\td E<0$ are obtained by replacing $q$ and $\td E$ by $|q|$ and $|\td E|$ in $\W $.}
\be
 -{\pi k M^2\over 2q\td E} -{\pi k\w^2 \td E^2\over 2q \td E}  \pm \pi k\w 
\sim -{\pi k M^2\over 2q\td E} \big(1+\eps ^2 \pm {2q\over M}\ \eps \big)
\label{lala}
\ee
The gravitational back reaction term, proportional to $\eps^2$, can be neglected provided $\eps\ll 1 $ and $\eps \ll q/M$, the latter being in general a stronger
condition since $M\geq q$. This requires $|\w_0|\ll q/\td E$.
%
In this limit we have
\bea
{\cal W} 
&\cong & {{\cal T}V_{d-2}\over (2\pi)^{d-1}} \sum_{k=1}^\infty  {(-1)^{k+1}\over 2qE } \left( {2q\td E\over k} \right)^{d\over 2}
\int_{-|\w_0|}^0 d\omega\  
 e^{-{\pi k M^2\over 2q\td E}} \big( e^{-\pi k\w}-e^{\pi k\w} \big) 
\nn\\
&\cong &  {4{\cal T}V_{d-2}\over (2\pi)^{d} } \sum_{k=1}^\infty {(-1)^{k+1}\over k^2} \left( {2q\td E\over k} \right)^{d-2\over 2} 
e^{-{\pi k M^2\over 2q\td E}}  \big(\cosh(\pi k\w_0) -1 \big)  
\label{keb}
\eea  
Note that for these modes with $|\w|<|\w_0|$  it is $M^2$, rather than $M^2-q^2$, what appears in the exponent.

In the Rindler case (\ref{gabo}),  restricting the integrations to the same modes with $|\w|<|\w_0 | \ll M^2/eE$, one obtains
\be
{\cal W} 
= {{\cal T}V_{d-2}\over (2\pi)^{d-1}} \sum_{k=1}^\infty  {(-1)^{k+1}\over 2qE } \left( {2q\td E\over k} \right)^{d\over 2}
\int_{-|\w_0|}^0 d\omega\  
 e^{-{\pi k M^2\over 2q\td E}} \big( 1 -e^{2\pi k\w} \big) 
\label{unamas}
\ee
In general, this is different from (\ref{keb}), as expected, since the models are different. Nevertheless, it is interesting that it gives the same
result as (\ref{keb}) when the integral is restricted to low energy modes, $\w_0\ll 1$, i.e. 
the probabilities agree precisely in the regime where the rate is significant.
 To see this, we note that, because of the exponential factor, $\W $ in (\ref{keb})  is very small unless ${M^2\over q\td E} < O(1)$. Since
  $\w\ll {q\over\td E} \leq {M^2\over q\td E}$,  this condition implies $\w_0  \ll 1 $, i.e. low-frequency modes. 
 In this case we can assume $k\w_0\ll 1 $ and expand the $\cosh(\pi k \w_0 )$.\footnote{Since the series is convergent, large values of $k$ with $k > O(1/\w_0)$
give subleading contributions.} We find
\be
{\cal W} =  {\pi {\cal T}V_{d-2}\over  (2\pi)^{d-1} } \sum_{k=1}^\infty (-1)^{k+1} \left( {2q\td E\over k} \right)^{d-2\over 2} 
e^{-{\pi k M^2\over 2q\td E}}  \ \w_0^2\ ,
\label{kgb}
\ee 
which exactly agrees with (\ref{unamas}) in the same limit.

\medskip

The formula (\ref{keb}) also clarifies the apparent puzzle mentioned above concerning the 
singular behavior of eq. (\ref{greed}) for $d=2$ at zero $\td E$. 
This  divergence originates from modes with large $\w $, for which the gravitational
back reaction term $ \w^2 \td E^2$ is important. In restricting the integration to modes
for which the gravitational back reaction is small, we find (\ref{keb}) which has a regular zero field limit in any dimensions, as expected.

\medskip

In the present case it is not clear how to disentangle the Unruh effect from the Schwinger process, 
since there is no infinite radial volume factor. The Unruh effect should be significant near the horizon, whereas typically the Schwinger effect 
appears in the region where the electric field has a non-vanishing value, giving a rate typically proportional to the volume of this region.
We recall that the geometry (\ref{Bmodel}) of the present model is non-trivial. The pair creation process arises as the net result
of this complicated combination of Rindler horizon with gauge fields and back reaction.
It is possible that it makes no sense to attempt to separate Unruh and Schwinger effects  at any value of the electric field.





\section{Bosonic string model}

The partition function for the two-parameter $(b,\tilde b)$ magnetic model  (\ref{kkm}) in bosonic string theory was found  in \cite{exactly}. 
The models with $(b,0)$ and $(0,\tilde b)$, being related by T-duality, have identical partition function with the exchange of $R\to \a'/R$. 
Here we will set $\tilde b=0$ and use units where $\a'=1$. One gets
\bea
&&Z = 
c_1 \int_{\cal F} {d^2\tau\over  \tau_2^{13}} \ e^{4\pi\tau_2} |f(e^{2i\pi \tau})|^{-48}
\sum_{(w',m)\neq(0,0)} \exp (-{\pi R^2\over \tau_2} |w'-\tau m|^2)
\nn\\
&& \times\ e^{-{\pi (\chi-\bar\chi)^2\over 2\tau_2}} { 1\over \big| \sin(\pi\chi)\big|^2}\prod_{r=1}^\infty 
{  \big| (1-e^{2\pi i r\tau } )\big|^4 \over  
\big| (1-e^{2\pi i ( r\tau +\chi )} )(1-e^{2\pi i ( r\tau -\chi )} )\big|^2}
\label{ZFbos}
\eea
where 
\be
\chi = bR (w'-\tau m) \ ,\qquad \bar \chi = bR (w'-\bar \tau m) \ ,\qquad b>0\ .
\label{chies}
\ee
\be
f(e^{2i\pi \tau})=\prod_{r=1}^\infty (1-e^{2i r\pi \tau})\ ,\qquad c_1= {V_{D-3}R\over 4(2 \pi)^{D-3}}\ ,\ \ D=26\ .
\ee
Here ${\cal F}$ denotes as usual the fundamental domain of $SL(2,Z)$ defined by ${\cal F}=\{| \tau_1|\leq 1/2, \ |\tau|^2\geq 1\}$ in the upper half-plane $\tau_2>0$. The partition function (\ref{ZFbos}) has a divergent zero field limit corresponding to the area of the plane 
$(x_1,x_2)\equiv (r,\varphi )$.
Alternatively, one could project out the factor coming from the integral of the constant modes of $x_1,x_2$ (this prescription was adopted in
 \cite{exactly}). This leads to (\ref{ZFbos}) with an additional factor $\big| \chi \big|^2/\tau_2$. In both cases the partition function is modular invariant.

Before discussing some features of the general structure of this partition function, it is useful to study it in the particle limit
and compare with the results of the previous section.
In the T-dual language that we are studying, with $\tilde b=0$ and $b\neq 0$, this Kaluza-Klein particle corresponds to a winding state.  

\subsection{Schwinger effect for winding string states}

In order to obtain the partition function for the  winding state from (\ref{ZFbos}), the first step is to drop the factors $|f( e^{2i\pi \tau} )|$ 
and the factors in the product from $r=1$ to $\infty $ associated with
string excitations. 
Next, by a Poisson resummation in $w'$, i.e. by using the formula
\be
\sum_{w'} F(w') = \sum_n \int_{-\infty}^\infty d\mu \ e^{2\pi i n \mu } F(\mu )
\label{poip}
\ee
we go to the Hamiltonian representation where the physical states are exhibited explicitly.
We are interested in states with $m\neq 0$ and $n=0$. 
Also, in the field theory limit, the  integral over the fundamental domain region is to be extended 
to the full strip $|\tau_1| \leq {1\over 2}$, $\tau_2>0$. By a change of integration variable, $x=\mu- m\tau_1$, 
the integration over $\tau_1$ is trivial (equal to 1) and we   finally obtain ($s\equiv \tau_2$)
\be
Z_{\rm particle}  = 2c_1\int_0^\infty {ds\over  s^{13}} \int_{-\infty}^\infty  dx \  e^{-{\pi R^2x^2\over s}}
 \sum_{m=1}^\infty  e^{-\pi s(R^2   m^2- 2 b^2 R^2 m^2 -4)}
 { 1 \over  \big|\sin(\pi bR(x-i m s)) \big|^2 }
\label{Zparti}
\ee
The terms $- 2 b^2 R^2 m^2 -4$ in the exponent  come from normal ordering of the Hamiltonian, including all string oscillators \cite{exactly}. Therefore  they will be
dropped in what follows, since they do  not arise in the particle theory (these terms are absent in the supersymmetric theory, see \cite{RT2} and below).

To make contact with the spectrum (\ref{jamil}), we have to expand the sine function. We get
\be
Z_{\rm particle}  = 8 c_1\int_0^\infty {ds\over  s^{13}} \int_{-\infty}^\infty dx\  e^{-{\pi R^2 x^2\over s}}
\sum_{m=1}^\infty e ^{-\pi s R^2   m^2} 
\sum_{l_L,l_R=0} ^\infty e^{ - 2\pi s mbR(l_L +l_R+1)}   e^{2\pi i x bR(l_L-l_R) }
\label{Zsum}
\ee
 Integrating over $x$, we find
 \be
Z_{\rm particle}  = {8c_1\over  R} \sum_{m=1}^\infty
\int_0^\infty {ds\over  s^{25/2}} 
  \sum_{l_L,l_R=0} ^\infty  e^{-\pi s(R^2   m^2+b^2  (l_L-l_R)^2 )}  e^{ - 2\pi s mbR(l_L +l_R+1)}  
\label{Zspec}
\ee
This is what one would find from a Hamiltonian
$$
H= 2m b R(l_L +l_R+1)+ R^2   m^2+b^2  (l_L-l_R)^2+p_\mu^2, \qquad \mu=0,...,D-4\ ,\ \ D=26\ . 
$$
Changing $b\to \td b $, $R\to 1/R$ and $m\to n$, this is exactly the Hamiltonian (\ref{jamil}).
\medskip

Recalling 
\be
l=l_L-l_R\ ,\qquad l_L+l_R =2k_r +|l|\ ,\qquad k_r=0,1,2,...., 
\ee 
we write
\be
\sum_{l_L,l_R=0} ^\infty = \sum_{l=0}^\infty \sum_{k_r=0}^\infty + \sum_{l=-\infty}^{-1} \sum_{k_r=1}^\infty
 \ee
Then we can write $Z_{\rm particle }= Z_++Z_-$ with
\bea
Z_+ &=& {8c_1\over  R} \sum_{m=1}^\infty\sum_{l=0}^\infty \sum_{k_r=0}^\infty 
\int_0^\infty {ds\over  s^{25/2}}\ 
\ e^{-\pi s(R^2   m^2+b^2 l^2 )} e^{ - 2\pi s mbR(2k_r +l+1)}  
\nn\\
Z_- &=& {8c_1\over  R} \sum_{m=1}^\infty \sum_{l=-\infty}^{-1} \sum_{k_r=1}^\infty 
\int_0^\infty {ds\over  s^{25/2}}\ 
\ e^{-\pi s(R^2   m^2+b^2 l^2 )} e^{ - 2\pi s mbR(2k_r -l+1)}  
\eea
Summing over $k_r$, we find
\bea
Z_+ &=& {4c_1\over  R}  \sum_{m=1}^\infty\sum_{l=0}^\infty 
\int_0^\infty {ds\over  s^{25/2}}\  
\ e^{-\pi s(R^2   m^2+b^2 l^2 )} e^{ - 2\pi s mbR l}  {1\over \sinh (2\pi s mbR  ) }
\nn\\
Z_- &=& {4c_1\over  R}  \sum_{m=1}^\infty \sum_{l=-\infty}^{-1} 
\int_0^\infty {ds\over  s^{25/2}}\ 
\ e^{-\pi s(R^2   m^2+b^2 l^2 )} e^{  2\pi s mbR l}   {1\over \sinh (2\pi s mbR  ) }
\eea
Now consider the analytic continuation to the Lorentzian electric field configuration. This requires an additional prescription
which is not supplied by  the Euclidean magnetic partition function.
In section 3 we have seen that, in going to the Lorentzian electric field configuration $b\to iE$, 
at the same time  we have to change $l\to i\w $, and $\w $ is a continuous variable
representing the Rindler energy of a given mode.
One has to be careful in the Wick rotation $l\to i\w $, so we will proceed in two steps.
We first replace $l$ by a continuous variable $\ell $. We have 
\bea
Z_+ &=& {\T\over 2\pi } {4c_1\over  R}  \sum_{m=1}^\infty \int_{0}^\infty d\ell 
\int_0^\infty {ds\over  s^{25/2}}\ 
\ e^{-\pi s(R^2   m^2+b^2 \ell^2 )} e^{ - 2\pi s mbR \ell}  {1\over \sinh (2\pi s mbR  ) }
\nn\\
Z_- &=& {\T\over 2\pi }  {4c_1\over  R} \sum_{m=1}^\infty \int_{-\infty}^0 d\ell
\int_0^\infty {ds\over  s^{25/2}}\  
\ e^{-\pi s(R^2   m^2+b^2 \ell^2 )} e^{  2\pi s mbR \ell }   {1\over \sinh (2\pi s mbR  ) }
\eea
We have used the same normalization as before for the density of states. 
The Wick rotation $\ell \to i\w $, for both regions of integrations $\ell>0$ and $\ell <0$ must be done
to the negative imaginary axes for $\w $, since, as discussed in the quantum field theory treatment,  the negative frequency modes are those
contributing to the rate in the quantization in the right quadrant. Starting with clockwise oriented contours,
 there will be a relative sign between $Z_+$ and $Z_-$ contributions because in one case one has to reverse
the sense of integration of $\w $. The result is
\bea
Z_+ &=& {2\T c_1\over \pi R}  \sum_{m=1}^\infty \int_{-\infty}^0 d\w
\int_0^\infty {ds\over  s^{25/2}}\ 
\ e^{-\pi s(R^2   m^2+E^2 \w^2 )} e^{  2\pi s mER \w }  {1\over \sin (2\pi s mER  ) }
\nn\\
Z_- &=& -{2\T c_1\over \pi R}  \sum_{m=1}^\infty \int_{-\infty}^0 d\w
\int_0^\infty {ds\over  s^{25/2}}\  
\ e^{-\pi s(R^2   m^2+E^2 \w^2 )} e^{ - 2\pi s mER \w }   {1\over \sin (2\pi s mER  ) }
\eea
The pair creation rate  is equal to   $2{\rm Im} (Z_{\rm particle})$.
The imaginary part arises from the poles of ${1/ \sin (2\pi s m ER ) }$. As dictated by the definition of the Feynman propagator,
the integration contour passes by the right of the poles, and the imaginary part
is then given by $\pi$ times the residue at the poles,
\be
s_0= {k\over 2mER}\ ,\qquad {\rm Res}_{s_0}{1\over \sin (2\pi s m ER ) }={(-1)^k\over 2\pi m E R}\ .
\ee 
Thus we find
\bea
Z_+ &=&  {\T V_{d-2}\over (2\pi)^{d-1} } \sum_{m=1}^\infty \sum_{k=1}^\infty \int_{-\infty}^0 d\w \  {(-1)^k \over 2 m E R}  \left({2mER \over k}\right)^{25\over 2}\  
e^{- {\pi k\over 2mER} (M^2+\w^2 E ^2 )}e^ {\pi k \w }
\nn\\
Z_- &=&  - {\T V_{d-2}\over (2\pi)^{d-1} }   \sum_{m=1}^\infty  \sum_{k=1}^\infty \int_{-\infty}^0  
d\w \  {(-1)^k \over 2 m E R} \left({2mER \over k}\right)^{25\over 2}\ e^{- {\pi k\over 2mER} (M^2+\w^2 E ^2)} e^ {-\pi k \w }
\label{ZsW}
\eea
with $M=mR$ and $d=D-1=25$.
This exactly reproduces the result (\ref{derek}) found  directly from the differential equation in the previous section
(recall the T-dual dictionary: $E\to \td E$, $m\to n$ and $R\to 1/R$).
The integral over $\w $ can now be performed with identical results.

\subsection{Comments on the extension to general string states}

We  now comment on the computation of the pair production rate for {\it excited} winding string states.
In general, one expects that they should be pair produced since they are electrically charged as long as they have non-vanishing winding number.
In sect. 4.1 we have seen that obtaining the electric model from the magnetic model by analytic continuation
involves  a number of subtleties. In particular, the orbital angular momentum $l$ becomes a continuous variable
representing the  energy of the modes. Finding the correct prescription for the complete partition function involves many new issues. 
Here we outline  some aspects of the calculation.
We start with  (\ref{ZFbos}) and
perform Poisson resummation as in (\ref{poip}) to put the magnetic model in a suitable form
for analytic continuation. Expand each of the factors in  (\ref{ZFbos}),
\be
{1\over 1-e^{2\pi i ( r\tau \pm \chi )}} =\sum_{N_r^\pm =0}^\infty e^{2\pi i N_r^\pm ( r\tau \pm \chi )}\ ,\qquad \chi =bR(x-i\tau_2 m)\ ,
\ee
with $x=\mu-m\tau_1$. We obtain
\bea
&&Z = 
c_1 \sum_{\rm states} d_{\rm state}\int_{\cal F} {d^2\tau\over  \tau_2^{13}} \ e^{2\pi i (N_R-N_L-mw)\tau_1} \int_{-\infty}^\infty  dx \  e^{-{\pi R^2x^2\over \tau_2}+2\pi i n x} 
\nn\\
&&\times \ e^{-\pi\tau_2 (2N_R+2N_L +m^2R^2 - 2  m b R (J_R-J_L) - 2b^2 R^2 m^2-4) } \ e^{2\pi i x b R(J_R+J_L)}
\label{Zzz}
\eea
where $d_{\rm state}$ represents the degeneracy and
$$
N_{R,L}=\sum_{r=1}^\infty r (N_{r}^+ + N^-_{r})_{R,L}\ ,\qquad   J_{R,L}=\mp \big( l_{R,L}+{1\over 2}\big)+\sum_{r=1}^\infty  (N_{r}^+ - N^-_{r})_{R,L} \ 
$$
Next,  integrate over $x$ and get
\be
Z = 
{c_1\over R} \sum_{\rm states} d_{\rm state} \int_{\cal F} {d^2\tau\over  \tau_2^{25/2}} \ e^{2\pi i (N_R-N_L-mw)\tau_1} 
 \ e^{-\pi\tau_2 M^2 } 
\label{Zhh}
\ee
where 
\be
 M^2 = 2N_R+2N_L + m^2 R^2+ ({n\over R} + b(J_R+J_L))^2 -2  m b R (J_R-J_L) - 2b^2 R^2 m^2-4 
\ee
This of course reproduces the mass spectrum found in \cite{exactly}.
The last two terms $- 2b^2 R^2 m^2-4 $ imply the presence of  the usual bosonic string tachyon. 
To avoid this problem, one can start with the superstring partition function \cite{RT2}
\bea
&&Z_{\rm susy} = c
\int_{\cal F} {d^2\tau\over  \tau_2^5} \sum_{(w',w)\neq(0,0)} \exp (-{\pi R^2\over \tau_2} |w'-\tau w|^2) \big| f(e^{2i\pi \tau})\big|^{-12}
\nn\\
&& \times { \big|\sin(\pi\chi/2)\big|^8\over \big| \sin(\pi\chi)\big|^2}\prod_{n=1}^\infty 
{  \big| (1-e^{2\pi i ( n\tau +{\chi \over 2})} )(1-e^{2\pi i ( n\tau -{\chi\over 2})} )\big|^8 \over  
\big| (1-e^{2\pi i ( n\tau +\chi )} )(1-e^{2\pi i ( n\tau -\chi )} )\big|^2}
\label{ZF}
\eea
with $\chi $ defined in (\ref{chies}).
Proceeding in the same way,
one finds a similar formula as (\ref{Zhh}), with a different degeneracy $d_{\rm state}^{\rm susy}$, and with 
\be
 M^2_{\rm susy} = 2  N_R+2 N_L + m^2 R^2+ ({n\over R} + b(\hat J_R+\hat J_L))^2 -2  m b R (\hat J_R-\hat J_L)
\ee
Now the spectrum is tachyon free below some critical value of the magnetic field \cite{RT2,magnetic} and
the angular momentum operators $\hat J_R, \ \hat J_L$ take  both integer and half-integer values 
(while $N_R,\ N_L =0,1,2,...$).

The problem is how to implement analytic continuation to the electric field model in Lorentzian space.
This seems to require $b\to iE$ and  $J_R+J_L\to i (J_R+J_L)$, 
since the total angular momentum is the variable conjugate to $\varphi$ and $\varphi \to -i t$.
From sect. 4.1, we know that this is the case for the zero mode part where $l\to i\w $.
It should be the case that the spectra of $\hat J_R+ \hat J_L$ and $\hat J_R- \hat J_L $ operators contain  continuous and  discrete
parts related to Rindler energies and radial modes.

One could attempt to directly quantize strings
in the Lorentzian electric field configuration. This also involves a number of subtle issues (see \cite{Costa,Pioline,Berkooz} for discussions).
In the magnetic model, the zero modes satisfy the  algebra of creation and annihilation operators. There is an obvious definition of Fock space and
a physical string spectrum which is easily understood: the magnetic field only introduces corrections to the mass of each string state of given quantum numbers \cite{RT2}.
In the  electric field configuration, the zero modes satisfy a Heisenberg algebra and it is not clear how the Fock space should be defined in
order to reproduce the rates found in the previous sections. 

Alternatively, one may attempt to directly compute  the Euclidean path integral using the conformal $\sigma$-model associated with the background (\ref{Amodel}), i.e. the Euclidean version of the $E$-model (\ref{Amodel}). This 
leads to the partition function (\ref{ZFbos}) or (\ref{ZF}) with the change $b\to i E$.
This partition function with $b\to iE$, and without any additional change,  does not appear to be the proper starting point to obtain the pair production rate. 
Such partition function (which appeared in the literature in the context of cosmological Milne  universes  in \cite{Costa,Pioline,Berkooz}),
presents a number of features which are not well understood. It has   an infinite number of poles 
in the interior of the fundamental domain that lead to logarithmic divergences.
These poles can be more conveniently visualized on the strip $|\tau_1| \leq {1\over 2}$, $\tau_2>0$ by unfolding the fundamental domain: 
the pair $(w',m)$ can be written  as  $(w',m)=k(p,q)$ where $k$ is an integer and $p,\ q$ are relatively primes. Then, by a modular transformation, one can set $p$ or $q$ to zero and
 the sum over $(p,q)$ can be traded by a sum over copies of the fundamental domain, finally obtaining a total integration region given by the strip 
 $|\tau_1| \leq {1\over 2}$, $\tau_2>0$.
 A similar situation arose for the first time in a  different context \cite{malda}: strings propagating in 
thermal  $AdS_3$ backgrounds (or $H_3/Z$), representing the Euclidean BTZ black hole. Indeed, the partition function of \cite{malda} 
is very similar  to the bosonic string partition function (\ref{ZFbos}).
In that case, the divergences were interpreted as an infinite volume factor due to the fact that the long strings of $AdS_3$ feel a flat potential and can move to any radial position. 
For our present problem of  strings in the space (\ref{Amodel}), it is not clear how the partition function (\ref{ZFbos}) with  $b\to i E$ could capture
the Lorentzian physics, since in the Euclidean approach the time variable is  treated as an angle and the conjugate quantum numbers (such as
$l_{L,R}, S_{L,R}$) are discrete, while, as discussed above, a continuous part is expected. 
Clearly, it would be interesting to obtain a closed formula for the pair creation rate for the complete  string theory spectrum.

\begin{acknowledgments}
We acknowledge support by  MCYT FPA 2007-66665.
\end{acknowledgments}

\end{document}